\documentclass[twocolumn,showpacs,preprintnumbers,amsmath,amssymb]{revtex4-1}

\usepackage{graphicx}
\usepackage{dcolumn}
\usepackage{bm}
\usepackage{color}

\begin{document}

\title{Ultra-coherent single photon source}

\author{H. S. Nguyen$^1$}
\author{G. Sallen$^1$}
\altaffiliation[Present address: ]{Laboratoire de Physique et Chimie des Nano-Objets, 135 avenue de Rangueil, 31077 Toulouse, France.}
\author{C. Voisin$^1$}
\author{Ph. Roussignol$^1$}
\author{C. Diederichs$^1$}
\email{carole.diederichs@lpa.ens.fr}
\author{G. Cassabois$^{1,2,3}$}
\affiliation{$^1$Laboratoire Pierre Aigrain, Ecole Normale Sup\'erieure, CNRS (UMR 8551),
Universit\'{e} P. et M. Curie, Universit\'{e} D. Diderot, 24, rue Lhomond, 75231 Paris Cedex 05, France}
\affiliation{$^2$Universit\'e Montpellier 2, Laboratoire Charles Coulomb UMR5221, F-34095, Montpellier, France}
\affiliation{$^3$CNRS, Laboratoire Charles Coulomb UMR5221, F-34095, Montpellier, France}

\date{\today}

\begin{abstract}
We present a novel type of single photon source in solid state, based on the coherent laser light scattering by a single InAs quantum dot. We demonstrate that the coherence of the emitted single photons is tailored by the resonant excitation with a spectral linewidth below the radiative limit. Our ultra-coherent source opens the way for integrated quantum devices dedicated to the generation of single photons with high degrees of indistinguishability.
\end{abstract}

\maketitle

Semiconductor quantum dots (QDs) are promising candidates for the realization of reliable single photon emission devices in the contexts of fundamental quantum optics experiments or quantum information applications. In the pioneering work of Michler \emph{et al.}, intensity correlation measurements on the photoluminescence signal of single QDs brought the first experimental evidence of the quantum nature of the QD radiation \cite{Michler:00}. Subsequent experiments demonstrated the possibility to use single QDs for the generation of polarization-entangled single photon pairs \cite{Stevenson:06,Akopian:06,Dousse:10}, or indistinguishable single photons \cite{Santori:02,Ates:09}. An important general issue is to reach the so-called fundamental radiative limit $T_2=2T_1$, with $T_1$ the population lifetime and $T_2$ the coherence lifetime, where the linewidth of the photoluminescence spectrum corresponds to its lower limit $2\hbar/T_2=\hbar/T_1$. Several strategies have been followed in that sense such as the shortening of population lifetime induced by the Purcell effect \cite{Santori:02,Varoutsis:05}, or the reduction of pure dephasing by resonant excitation \cite{Muller:07,Ates:09}. To the best of our knowledge, all the measurements reported in the literature focus on the incoherent photoluminescence signal for achieving single photon emission.

In this letter, we present a novel type of single photon source based on the coherent laser light scattering by a single semiconductor QD. Besides the anti-bunching effect showing the non-classical nature of the emitted light, we observe that the QD emission spectrum is determined by the spectrum of the resonant excitation laser. This intrinsic striking feature of our system demonstrates the generation of single photons within a spectral linewidth below the radiative limit, resulting in a so-called ultra-coherent character. Our study brings the first evidence in a solid state system of the peculiar physical phenomenon where the radiated electric field has the classical character of the resonant excitation source while the emission statistics exhibit the quantum nature inherited from the light-matter coupling at the scale of a single system. Our ultra-coherent single photon source opens the way for integrated quantum device where the generation of indistinguishable single photons is tailored by the excitation laser source.

Resonant excitation experiments on single QDs were performed in an orthogonal excitation-detection geometry \cite{Muller:07}. The sample consists of a low density layer (0.01~$\mu m^{-2}$) of self-assembled InAs/GaAs QDs embedded in a planar $\lambda_0$-GaAs microcavity formed by two distributed-Bragg reflectors. The top and bottom Bragg mirrors are respectively made of 11 and 24 pairs of $\lambda_0/4$ layers of AlAs and AlGaAs which leads to a microcavity quality factor of 2500. Resonant excitation of the QD is performed by a tunable cw external cavity diode laser. A Michelson interferometer allows high-resolution Fourier transform spectroscopy of the QD resonant emission (RE), where the first-order correlation function $g^{(1)}(\tau)$ is deduced from the contrast of the interferences \cite{Berthelot:06}. A Hanbury-Brown and Twiss (HBT) setup is also used to perform intensity correlation measurements under resonant excitation \cite{Michler:00} and to measure the second-order correlation function $g^{(2)}(\tau)$.

\begin{figure}[htb]
\begin{center}
\includegraphics[width=8.5cm]{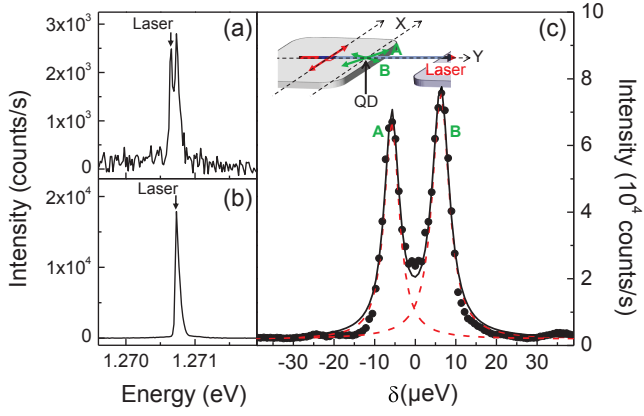}
\end{center}
\caption{\label{RE}{Micro-photoluminescence spectrum for a laser detuning with the exciton energy $\delta=E_{L}-E_{X}$ of -74~$\mu$eV (a) and -6~$\mu$eV (b). (c) Intensity of the QD resonant emission as a function of $\delta$. The solid line is the sum of two Lorentzian fits (dashed lines) of width 4.4~$\mu$eV and 5.0~$\mu$eV, separated by 12~$\mu$eV, corresponding to the two linear orthogonal polarizations A and B, of the neutral exciton. Inset: Scheme of the tilted QD axis A and B with respect to the main crystallographic axis $[110]$ and $[1\overline{1}0]$, labeled as X and Y.}}
\end{figure}

Figures~\ref{RE}(a) and \ref{RE}(b) show micro-photoluminescence ($\mu$PL) spectroscopy on a single QD at 10K. Figure~\ref{RE}(a) is a quasi-resonant $\mu$PL spectrum, where the energy of the resonant diode laser $E_L$ is slightly detuned from the mean exciton energy $E_X$ ($\delta=E_{L}-E_{X}=-74~\mu eV$). We observe that the laser light scattering is highly reduced in the orthogonal excitation-detection setup and it is here comparable to the intensity of the excitonic emission. When the laser is strictly resonant with an exciton line in figure~\ref{RE}(b), the intensity is enhanced by a factor 60 due to the resonant excitation and this enhancement is here only limited by the QD saturation properties. Figure~\ref{RE}(c) is the RE spectrum of the same single QD where the intensity of the exciton emission under resonant excitation is plotted as a function of the laser detuning $\delta$. In comparison to the previous $\mu$PL spectrum where the exciton linewidth is not resolved, we now observe two Lorentzian lines of width 4.4~$\mu$eV and 5.0~$\mu$eV, separated by 12~$\mu$eV. Polarization-resolved measurements show that they correspond to two linear orthogonal polarizations, hereafter called A and B, as expected for the fine structure splitting of the neutral exciton \cite{Hogele:04}.

In the following, we investigate the different contributions in the RE signal. It is in fact known that the RE signal consists of the superposition of the coherent resonant Rayleigh scattering (RRS) and the incoherent resonant photoluminescence (RPL). The RRS component has a spectrum exactly corresponding to the one of the resonant excitation laser. The incoherent RPL has a more complex profile evolving from a single Lorentzian of linewidth $2\hbar/T_2$ to the so-called Mollow triplet on increasing the excitation power \cite{AtomePhoton}. Note that in figure~\ref{RE}(c), for a given detuning, the recorded intensity consists in the sum of these two components. In fact, one expects the RE signal to be dominated by the coherent RRS at low incident power, and on the contrary by incoherent RPL close to the QD saturation power. This latter regime has already been well studied in single QDs \cite{Muller:07,Melet:08,Ates:09,Flagg:09} with the evidence of the Mollow triplet \cite{Flagg:09}, characteristics of this peculiar high power regime.

\begin{figure}[htb]
\begin{center}
\includegraphics[width=8.5cm]{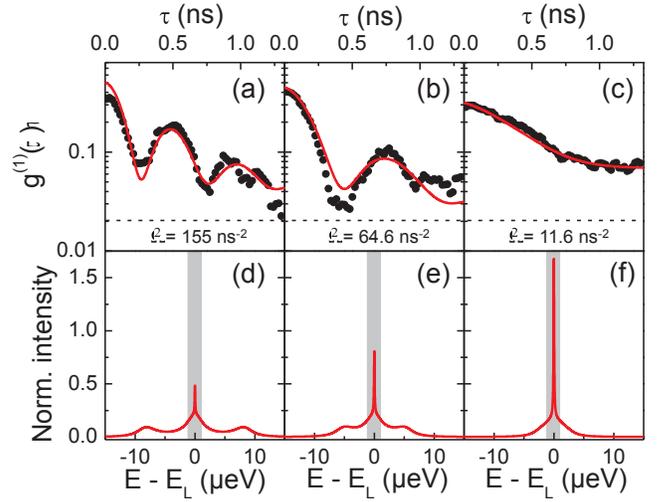}
\end{center}
\caption{\label{g1}{(a), (b) and (c) High-resolution Fourier transform spectroscopy of the resonant emission of the exciton line B, at $\delta=6$~$\mu$eV, for decreasing powers corresponding respectively to $\Omega^2=155$~ns$^{-2}$, 64.6~ns$^{-2}$ and 11.6~ns$^{-2}$. Fit (solid line) to the first order correlation function $g^{(1)}(\tau)$ with $T_1=320$~ps, $T_2=410$~ps and $T_L=10$~ns. The background level is indicated with a dashed line. (d), (e) and (f) Corresponding calculated resonant emission spectrum deduced by inverse Fourier transform of the $g^{(1)}(\tau)$ function ($\Omega^2=155$~ns$^{-2}$, 64.6~ns$^{-2}$ and 11.6~ns$^{-2}$ respectively). The radiative-limit of width 2~$\mu$eV is schematized in grey and superimposed on each spectrum.}}
\end{figure}

Our study focuses on the low excitation regime determined by the coherent RRS. Figure~\ref{g1}(a, b \& c) shows high-resolution measurements by Fourier transform spectroscopy of the RE, for resonant excitation of the exciton line B ($\delta=6$~$\mu$eV), for three excitation powers corresponding to $\Omega^2=155$~ns$^{-2}$, 64.6~ns$^{-2}$ and 11.6~ns$^{-2}$, where $\Omega$ is the Rabi frequency. Our data are fitted (solid lines in Fig.~\ref{g1}(a, b \& c)) to the first-order correlation function $g^{(1)}(\tau)$ of the RE of a two-level system:
\begin{eqnarray}
g^{(1)}(\tau)&\propto&\text{e}^{-\frac{\tau}{T_L}} + A\left[\text{e}^{-\frac{\tau}{T_2}}+\text{e}^{-\alpha\tau}\left(B\cos(\beta\tau+\phi)\right)\right]
\end{eqnarray}
where $T_L$ is the coherence time of the resonant laser, $A$, $B$ and $\phi$ are three constants which depend on $\Omega$, $T_1$ and $T_2$, and $\alpha=\frac{1}{2}\left(\frac{1}{T_{1}}+\frac{1}{T_{2}}\right)$, $\beta=\sqrt{\Omega^2-\frac{1}{4}\left(\frac{1}{T_{1}}-\frac{1}{T_{2}}\right)^2}$. The first term in the $g^{(1)}(\tau)$ function corresponds to the coherent RRS whereas the second one is related to the incoherent RPL. The experimental data are fitted by taking for $T_1$ and $T_2$ the experimental values determined by independent experiments (not shown here). Time-resolved measurements of the RE signal intensity first lead to an exciton lifetime $T_1=320\pm20$~ps. Power-dependent experiments of the RE signal exhibit the standard power-broadening effect of a two-level system \cite{AtomePhoton} from which we deduce a decoherence time $T_2=410\pm10$~ps. As seen in Fig.~\ref{g1}(a, b \& c), with $T_1=320$~ps and $T_2=410$~ps, we fairly reproduce our experimental data. In particular, we account for the smooth transition from a non-monotonous decay corresponding to the Mollow triplet in the incoherent RPL regime, to a bi-exponential one where the long time component is determined by the coherence time $T_L$ of the resonant laser in the coherent RRS regime.

The inverse Fourier transform of the theoretical $g^{(1)}(\tau)$ function allows a direct comparison, in the spectral domain, of the different RE emission profiles (Fig.~\ref{g1}(d, e \& f)). The radiative-limit $\hbar/T_1$ of 2~$\mu$eV is also displayed on each spectrum as a guide of the eye of rectangular form. At high power (Fig.~\ref{g1}(d)), the incoherent RPL consists in a well-developed Mollow triplet with side bands separated by 8~$\mu$eV from the central peak of width 3~$\mu$eV. The RRS signal appears as a superimposed weak line, much narrower than the radiative limit. Upon reducing the resonant excitation power (Fig.~\ref{g1}(e \& f)), the Mollow triplet shrinks to a single Lorentzian line with sidebands, while the relative intensity of the narrow RRS peak increases and tends to overwhelm the RE spectrum.

\begin{figure}[htb]
\begin{center}
\includegraphics[width=7cm]{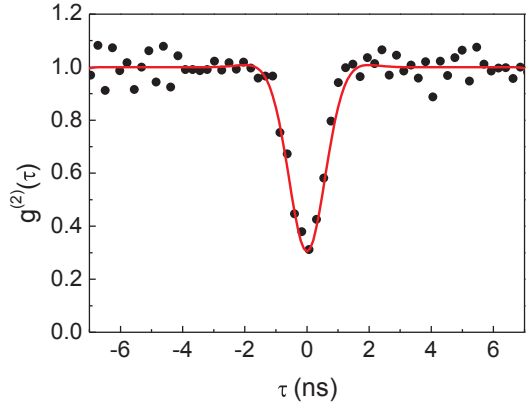}
\end{center}
\caption{\label{g2}{Intensity correlation measurements of the RE under resonant excitation of the exciton line B, for a power corresponding to $\Omega^2=2$~ns$^{-2}$. Fit (solid line) to the second order correlation function $g^{(2)}(\tau)$ with $T_1=320$~ps and $T_2=410$~ps.}}
\end{figure}

An important question is the existence of an upper limit for the RRS fraction and the corresponding physical parameters conditioning it. We have theoretically investigated the spectral distribution of the coherent and incoherent components of the RE signal by considering the optical Bloch rate equations. One can deduce the intensity of these two components under resonant excitation, and further calculate the ration between the RRS contribution and the total RE signal (RRS+RPL), which is given by:
\begin{eqnarray}
\frac{I_{RRS}}{I_{RE}}=\frac{T_2}{2T_1}\times\frac{1}{1+\Omega^2 T_1T_2}
\end{eqnarray}
In the case of an ideal two-level system at the so-called radiative limit ($T_2=2T_1$), one can see that the RE signal is completely dominated by the RRS for excitation powers such as $\Omega^2 T_1T_2\ll1$. On the contrary, as soon as pure dephasing appears leading to $T_2<2T_1$, the highest fraction of RRS in the RE signal is given by $T_2/2T_1$. In our system where $T_2=1.3 T_1$ (which is very closed to the radiative-limit), the RRS fraction can intrinsically not exceed 65\% of the RE signal. In Fig.~\ref{g2}, we display the normalized intensity auto-correlation measurements by the HBT setup under resonant excitation of the exciton line B, for an excitation power corresponding to $\Omega^2=2$~ns$^{-2}$ ($I_{RRS}/I_{RE}$=50\%). After subtracting the noise contribution and taking into account the system response function of 400~ps, the experimental data are well fitted by the theoretical second-order correlation function $g^{(2)}(\tau)$ of the RE of a two-level system, given by $g^{(2)}(\tau)=1-\left(\cos(\beta\tau)+\frac{\alpha}{\beta}\sin(\beta\tau)\right)\text{e}^{-\alpha\tau}$ \cite{QO,Flagg:09}. In particular, for low excitation powers ($\Omega\ll\frac{1}{2}\left|\frac{1}{T_1}-\frac{1}{T_2}\right|$), the second order correlation function follows the simple dependence on $T_1$ and $T_2$, $g^{(2)}(\tau)=1-\frac{1}{T_2-T_1}\left(T_2\text{e}^{-2\tau/T_2}-T_1\text{e}^{-2\tau/T_1}\right)$. The pronounced dip at zero time delay ($g^{(2)}(0)=0.3$), which is only limited by the system time response, shows that our system is indeed a very efficient single photon source, as expected in the regime where the RE spectrum is dominated by the RRS component \cite{QO,Flagg:09}. Moreover, even at this uncommon low-power regime, the RE intensity reaches 30~kcounts/s which is comparable to the emission rate of other complex devices \cite{Dousse:10,Claudon:10}. The resonant excitation of a single QD thus leads to this very peculiar situation where the first-order correlation function of the radiated electric field has the classical character of the excitation laser, whereas the second-order correlation function has a quantum nature inherited from the light-matter interaction at the scale of a single QD \cite{Walther:98}. As a matter of fact, this system represents a novel type of quantum device with ultra-coherent single photons controlled by the linewidth of the resonant excitation source and the photon statistics controlled by the QD properties.

In conclusion, we have demonstrated that when the RE of a single QD is dominated by the coherent laser light scattering, anti-bunching is observed together with a narrow emission spectrum below the radiative limit. This implies that single QDs can produce single photons with a decoherence time that is not limited anymore by the QD electronic properties. This ultra-coherent single photon source promises high degrees of indistinguishability of the emitted photons which is a crucial requirement for quantum information applications.

This work was supported financially by the "Agence Nationale pour la Recherche" project CAFE (ANR-09-NANO-022).

\end{document}